\documentclass[letter]{aa}  

\usepackage{graphicx}
\usepackage{subfig}
\usepackage{txfonts}
\usepackage{color}
\usepackage{lscape}
\usepackage[colorlinks = true,
            linkcolor = blue,
            urlcolor  = blue,
            citecolor = blue,
            anchorcolor = blue]{hyperref}
\usepackage{multirow}
\usepackage{soul}
\usepackage{comment}

\newcommand{\lgalaxies}{\textsc{L-Galaxies2020}}
\newcommand{\logmass}{$\log$(M$_*$/M$_\odot$)}
\newcommand{\mstar}{\mbox{$M_{*}$}}
\newcommand{\msun}{\mbox{$M_{\odot}$}}

\newcommand{\htwo}{H$_2$}
\newcommand{\sigmasfr}{$\Sigma_{\rm SFR}$}
\newcommand{\sigmaMS}{$\Sigma_{\rm SFR}{\rm MS}$}

\definecolor{amethyst}{rgb}{0.8, 0.0, 0.0}

\begin{document}

   \title{Evolution of the star formation rate surface density main sequence.}
   \subtitle{Insights from a semi-analytic simulation since $z=12$.} 
   \titlerunning{\sigmaMS\ in SAM since $z$=12}
   
\author{Jakub Nadolny\inst{1,2} 
\and Michał J. Michałowski\inst{1,3} 
\and Massimiliano Parente\inst{4,5,6} 
\and Mart\'in Solar\inst{1} 
\and Przemysław Nowaczyk\inst{1} 
\and Oleh Ryzhov\inst{1}
\and Aleksandra Leśniewska\inst{1}}

   \institute{Astronomical Observatory Institute, Faculty of Physics and Astronomy, Adam Mickiewicz University, ul.~S{\l}oneczna 36, 60-286 Pozna{\'n}, Poland \\ \email{jakub.nadolny@amu.edu.pl,quba.nadolny@gmail.com} 
    \and Instituto de Astrof\'isica de Canarias, E-38205 La Laguna, Tenerife, Spain
    \and Institute for Astronomy, University of Edinburgh, Royal Observatory, Blackford Hill, Edinburgh, EH9 3HJ, UK 
    \and INAF, Osservatorio Astronomico di Trieste, via Tiepolo 11, I-34131, Trieste, Italy 
    \and SISSA, Via Bonomea 265, I-34136 Trieste, Italy 
    \and IFPU, Institute for Fundamental Physics of the Universe, Via Beirut 2, 34014 Trieste, Italy 
    }
             
   \date{Received ---; accepted ---}

  \abstract%
  {Recent high-redshift ($z>4$) spatially resolved observations with the James Webb Space Telescope have shown the evolution of the star formation rate (SFR) surface density (\sigmasfr) and its main sequence in the \sigmasfr-$M_*$ diagram (\sigmaMS). The \sigmaMS\ is already observed at cosmic morning ($z\sim7.5$). The use of \sigmasfr\ is physically motivated because it is normalized by the area in which the star formation occurs, and this indirectly considers the gas density. The \sigmasfr-$M_*$ diagram has been shown to complement the widely used (specific) SFR-$M_*$, particularly when selecting passive galaxies.}
  {We establish the \sigmasfr\ evolution since $z=12$ in the framework of the \lgalaxies\ semi-analytical model (SAM), and we interpret recent observations.}
  {We estimated \sigmasfr(-$M_*$) and the cosmic star formation rate density (CSFRD) for the simulated galaxy population and for the subsamples, which were divided into stellar mass bins in the given redshift.}
  {The simulated \sigmasfr\ decreases by $\sim3.5$ dex from $z=12$ to $z=0$. We show that galaxies with different stellar masses have different paths of \sigmasfr\ evolution. We find that \sigmaMS\ is already observed at $z\sim11$. The simulated \sigmaMS\ agrees with the observed one at $z=0, 1, 2, 5$, and $7.5$ and with individual galaxies at $z>10$. We show that the highest \sigmaMS\ slope of $0.709\pm0.005$ is at $z\sim3$ and decreases to $\sim0.085\pm0.003$ at $z=0$. This is mostly driven by a rapid decrease in SFR with an additional size increase for the most massive galaxies in this redshift range. This coincides with the dominance of the most massive galaxies in the CSFRD from the SAM. Observations show the same picture, in which the \sigmasfr\ evolutionary path depends on the stellar mass, that is, more massive galaxies have higher \sigmasfr\ at all redshifts. { Finally, using the slope and normalization evolution, we derived the simulated \sigmaMS\ as a function of stellar mass and redshift}.}
  {}


   \maketitle

\section{Introduction}
\label{sec:intro}
The rate at which a galaxy forms stars is one of the most fundamental parameters providing insight into the assembly history of cosmic mass. The Schmidt-Kennicutt law \citep{Schmidt1959ApJ...129..243S,Kennicutt1998ApJ...498..541K} shows a correlation of the star formation rate (SFR) with the density of the gas that is used to form stars, from which the SFR-stellar mass \mstar\ correlation follows. 
Many studies showed a tight sequence in the SFR-\mstar\ diagram that is called star formation main sequence (SFMS) in optically selected star-forming galaxies  \citep[SFGs; ][]{Brinchmann2004MNRAS.351.1151B,Noeske2007ApJ...660L..43N,Speagle2014ApJS..214...15S,Nadolny2023ApJ...952..125N}, in very low-mass SFGs \citep{Nadolny2020A&A...636A..84N}, or in infrared selected SFGs \citep{Michalowski2012A&A...541A..85M,Michalowski2017MNRAS.469..492M,Koprowski2014MNRAS.444..117K,Koprowski2016MNRAS.458.4321K,Otero2024Lockman}. Another way of expressing the star formation activity is via the specific SFR (sSFR), which is the SFR normalized by the stellar mass. The sSFR-\mstar\ relation indicates the recent level of SF (including quenched SF), and similarly to the SFMS, it also evolves with cosmic time. The overall cosmic SFR density \citep[CSFRD;][]{MadauDickinson2014,Zavala2021ApJ} has also been shown to evolve and to reach its maximum at $z\sim 2$, with a steady decay with increasing redshift {  \citep{Bouwens_2022,Donnan2024MNRAS}}. 

In recent years, another more physically motivated parameter related to SF emerged {  \citep{WuytsSigmaSFR2011,Salim2023ApJ...958..183S,Calabro2024arXiv240217829C}.} This is the SFR surface density, which is defined as $\Sigma_{\rm SFR}={\rm SFR}/(2\pi R_e^2)$[\msun/yr/kpc$^2$], where $R_e$ is the radius of a galaxy (half-light radius in this work). This parameter is more fundamental than other SF tracers because it is normalized by the extent of the star-forming region. 
\citet{Salim2023ApJ...958..183S} showed that \sigmasfr\ complements the sSFR, particularly when passive galaxies are selected (because a galaxy with a constant SFR decreases in sSFR by definition). Its relation to the molecular gas density and the effectiveness of stellar feedback \citep{Salim2023ApJ...958..183S} also has a much deeper physical justification. The \sigmasfr\ has been found to increase by almost four orders of magnitude between $z=0$ and 10 \citep{WuytsSigmaSFR2011,Calabro2024arXiv240217829C}. Similar to the SFMS, a tight relation (or main sequences) in the \sigmasfr-$M_*$ diagram was found for SFGs (\sigmaMS). Several works \citep{WuytsSigmaSFR2011,Salim2023ApJ...958..183S,Calabro2024arXiv240217829C} showed the increasing normalization of \sigmaMS. \citet{Salim2023ApJ...958..183S} reported an increasing slope at $z=1$ and $2$ caused by the fast bulge-mass build-up in the most massive galaxies, while \citet{Calabro2024arXiv240217829C} observed the \sigmaMS\ at $z\sim5$ and $7.5$. The \sigmasfr\ measurement of high-$z$ compact star-forming nitrogen-rich galaxy \citep{Schaerer2024A&A...687L..11S} suggests that it may be useful to select this rare galaxy population.

Recent works on magnetohydrodynamics simulations \citep[e.g.][]{Girma2024MNRAS.527.6779G} reproduced the observed \sigmasfr\ at the high spatial resolution (tens of parsecs) in striking exactitude. These simulations lacked the volume needed to provide a statistically significant evolution of \sigmasfr, however.

The successes of \lgalaxies\ semi-analytic models (SAMs) in predicting many observables \citep{HenriquesLGal2020,Parente2023MNRAS.tmp..881P,Parente24,Nadolny2024arXiv240616533N,Wang2024arXiv240807743W} showed that the combination of resolution and statistical completeness of these simulations are adequate for drawing important conclusions about the evolution of \sigmasfr. This is the first study of \sigmasfr\ and the \sigmaMS\ evolution of galaxies divided by their stellar mass using an SAM. 

We probed the evolution of \sigmasfr\ and \sigmaMS\ since cosmic dawn from $z=12$ to $z=0$ using SAM \lgalaxies, which places low-, intermediate, and (very) high-$z$ (up to 14) observations in perspective. In particular, we investigated the \sigmasfr\ evolution of galaxies divided into stellar mass bins, following the analysis of the \sigmaMS\ evolution. We explored the possible physical causes of the slope increase in \sigmaMS\ observed at $z$ between 4 and 1. {  We show that the stellar mass and SFR are the fundamental parameters that define the different evolutionary paths of the simulated galaxy population.}

This paper is organized as follows. Section \ref{sec:data} describes the data from the literature and the SAM. In Section \ref{sec:results} we present our results. Section \ref{sec:discussion} presents the discussion and conclusions of this work. Throughout this paper, we use a cosmological model with
$h$ = 0.673, $\Omega_\Lambda = 0.685$, $\Omega_m = 0.315$ \citep{Planck2014}, and the \citet{Chabrier2003PASP..115..763C} initial mass function (IMF), as assumed in the applied SAM.

\section{Data}
\label{sec:data}

\subsection{\lgalaxies\ semi-analytic model}
\label{sec:data_coparison_sam}
We used the \lgalaxies\ SAM presented in \citet{Parente2023MNRAS.tmp..881P}, which is an extended version of the latest public release of the Munich SAM for galaxy evolution \citep{Henriques2015,HenriquesLGal2020}. This SAM accounts for a detailed dust production and evolution, and it includes an updated treatment of disk instabilities.
The \lgalaxies\ model starts from the dark matter (DM) halo merger trees of the \textsc{Millennium} simulation \citep{Springel_Milleniu2005Natur.435..629S} and populates them with baryons according to a number of important physical processes. Among these are gas infall into DM halos, gas cooling, disk and bulge formation, star formation, merger-driven starbursts, chemical enrichment, energetic feedback from stars and supermassive black holes (SMBHs), and environmental processes such as ram pressure and tidal stripping. We refer to the supplementary material of \cite{HenriquesLGal2020} and to \citet{Parente2023MNRAS.tmp..881P} for a detailed description of this SAM. 

{We used \texttt{StellarHalfLightRadius} as the effective radius $R_e$. The implemented star formation law is based on the \htwo\
surface density \citep{Bigiel2011ApJ...730L..13B,Leroy2013AJ....146...19L} and a    \citet{Chabrier2003PASP..115..763C} IMF was adopted.}

\subsection{\lgalaxies\ sample selection}
\label{sec:data_coparison_sam_sample}

To analyze \sigmasfr\ and \sigmaMS\ from $z\sim12$ to $z=0$, we used the whole simulated galaxy population with \mstar$>10^8$\msun. This is a statistically significant sample at all redshifts (the sample size in each snapshot is given in Tab. \ref{tab:fittingSigmaSFR}). Additionally, we divided the galaxies into stellar masses bins: between 8 and 9, 9 and 10, and above 10 in log(\mstar/\msun). 

\subsection{Observational data}
\label{sec:data_coparison_obs}
We used a compilation of the high-$z$ JWST data from the CEERS \citep{Finkelstein2023ApJ...946L..13F} and GLASS \citep{Treu2022ApJ...935..110T} surveys presented in \citet{Calabro2024arXiv240217829C}; massive dusty star-forming galaxies from RUBIES \citep{deGraaff2024Rubies}; massive (\mstar$>10^{11}$\msun) extended red dots \citep{gentile2024arxive} from COSMOS-Web \citep{Casey2023ApJ...954...31C}; and the SFG  \citep{Lines2024arXiv240910963L} from the CRISTAL-ALMA survey (PI: Rodrigo Herrera-Camus). We also included low-$z$ mean values from the GAMA survey \citep{Smith2011,Driver2011,Driver2016,Baldry2018}, from the COSMOS survey \citep{Scoville2007ApJS..172....1S,Weaver2022Cosmos2020}, {  and from the low-mass [OII] emitters at $z=1.4$ \citep{OTELO_OII_2021A&A} from the OTELO survey \citep{Bongiovanni2019}.} We also included observations of intermediate-$z$ objects from \citet{Williams2021ApJ...908...54W} and \citet{Rupke2023ApJ}. Finally, we also used the \sigmaMS\ estimated by \citet{Salim2023ApJ...958..183S} at $z=0, 1$, and $2$. { Regarding very high-$z$ ($>10$), we show galaxies from JADES \citep{Eisenstein2023JADES}, in particular, GS-z14-0 \citep{carniani2024}, GHZ2 \citep{Castellano2024ApJ}, and GNz11 \citep{Tacchella_2023} at $z\sim 14, 12$, and $11$, respectively.}

We estimated \sigmasfr\ consistently (i.e., using the half-light radius $R_e$ and the SFR based on the same IMF) in the SAM, the GAMA survey, and in the data from \citet{Williams2021ApJ...908...54W},\citet{gentile2024arxive}, \citet{carniani2024} and \citet{Tacchella_2023}. The remaining works give estimates of \sigmasfr using the same IMF and size estimates (half-light radii $R_e$; see \citealt{Salim2023ApJ...958..183S} for a discussion of the differences between the use of $R_e$ and isophotal size). While the \sigmaMS\ normalization changes for different size estimates and the scatter increases for $R_e$, the slopes of the obtained \sigmaMS\ are consistent in a given redshift \citep{Salim2023ApJ...958..183S}. This does not influence our conclusions, however, because we consistently used the same size estimates in the simulated data and observational results from the literature. Finally, we used the same stellar mass bins as for the simulated sample. 


\section{Results}
\label{sec:results}

\begin{figure*}[ht!]
\includegraphics[width=0.8\textwidth,clip]{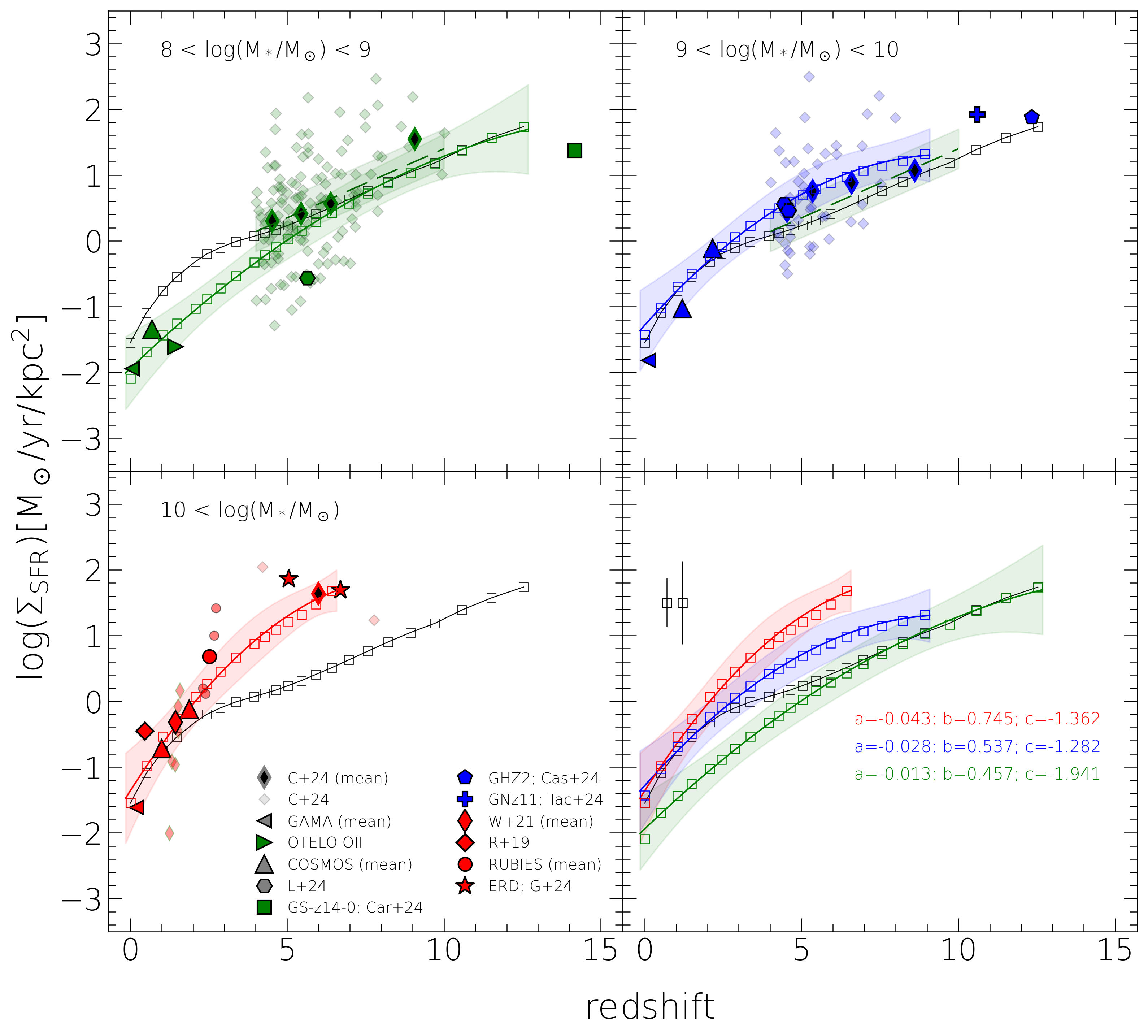}
\caption{Star formation rate surface density \sigmasfr\ as a function of redshift. The empty black squares show the median values for all the galaxies from \lgalaxies\ with \mstar$>10^8$\msun. {  The empty green, blue and red}squares show the median values for the \lgalaxies\ sample divided into stellar mass bins with log(\mstar/\msun) between 8 and 9, 9 and 10, and above 10. The lines of the same colors are the second-degree polynomial, and the coefficients are given in the bottom right panel. We show the observational data described in Section \ref{sec:data_coparison_obs}, color-coded by their stellar mass, indicated at the top of each panel. The \citet{Calabro2024arXiv240217829C} mean values are shown in bins of stellar mass and redshift, and individual galaxies are shown as small diamonds. The dashed green line (only in the top panels) shows the fit from \citet{Calabro2024arXiv240217829C}. The values of the median of \lgalaxies\ SAM are given in Table \ref{tab:data_pointsSigmaSFR}.
\label{fig:sigma_sfr}}
\end{figure*}


\subsection{SFR surface density in stellar mass bins}
\label{sec:results_mstarBins}
The evolution of \sigmasfr\ with the redshift of the \lgalaxies\ subsamples divided into stellar mass bins is shown in Figure \ref{fig:sigma_sfr} (bins with more than ten objects are shown). { The overall \sigmasfr\ decreases from $\sim60$ \msun/yr/kpc$^2$ at $z=12$ to $\sim0.05$ at $z=0$. The intermediate-mass and massive (blue and red squares) simulated galaxies follow a similar trend as the overall \lgalaxies\ sample (black squares) up to $z\sim4$. Above this redshift, the \sigmasfr\ of the most massive galaxies continue to increase and reach values of $\sim40$ \msun/yr/kpc$^2$ (at $z\sim7$); this is $\sim1$ dex above the overall galaxy sample. The intermediate-mass subsample is also above the overall sample ($\sim0.5$ dex) and shows a milder increase than the most massive sample. The low-mass galaxies show a constant decrease in \sigmasfr\ with cosmic time since $z=12$. They are the only subsample to reach beyond $z\sim9$. We show second-degree fits to the subsamples with the coefficients given in Figure \ref{fig:sigma_sfr}, and in Table \ref{tab:data_pointsSigmaSFR} we list the median values of \sigmasfr\ for three simulated stellar mass bins. 

The observational data from the literature are spread around the median values from the simulations. The fit from \citet{Calabro2024arXiv240217829C} runs parallel to (and between) the mean \lgalaxies\ points for low- and medium-mass galaxies. We note that virtually all (except for two) galaxies from \citet{Calabro2024arXiv240217829C} have stellar masses below $10^{10}$\msun. We further show the median values for their galaxies divided into stellar mass bins. These are found within the $2\sigma$ dispersion of the fitted function using simulated data for a particular stellar mass bin. This is also true when we consider individual galaxies (color-coded by the stellar mass). The spread is larger in this case, but the overall picture is that more massive galaxies have higher \sigmasfr throughout virtually the whole redshift range. }

\begin{table}[ht]
\begin{center}
	\caption{ $\log$(\sigmasfr)-redshfit evolution for the stellar mass bins shown in Figure \ref{fig:sigma_sfr}.  
\label{tab:data_pointsSigmaSFR}}
\begin{tabular}{ccccc}
$z$ & \multicolumn{4}{c}{$\log$(\sigmasfr) [\msun/yr/kpc$^2$]} \\\cline{2-5}
 & all & low \mstar & mid \mstar& high \mstar \\
\hline
0.0 & -1.55$\pm$0.63 & -2.09$\pm$0.47 & -1.43$\pm$0.44 & -1.54$\pm$0.48\\
0.5 & -1.09$\pm$0.55 & -1.69$\pm$0.36 & -1.02$\pm$0.42 & -0.98$\pm$0.38\\
1.0 & -0.76$\pm$0.47 & -1.43$\pm$0.27 & -0.69$\pm$0.43 & -0.53$\pm$0.35\\
1.5 & -0.54$\pm$0.39 & -1.25$\pm$0.23 & -0.49$\pm$0.38 & -0.26$\pm$0.31\\
2.1 & -0.32$\pm$0.3 & -1.03$\pm$0.18 & -0.23$\pm$0.36 & 0.07$\pm$0.29\\
2.4 & -0.2$\pm$0.28 & -0.88$\pm$0.17 & -0.09$\pm$0.36 & 0.27$\pm$0.27\\
2.9 & -0.1$\pm$0.26 & -0.72$\pm$0.18 & 0.06$\pm$0.35 & 0.46$\pm$0.27\\
3.4 & -0.01$\pm$0.24 & -0.53$\pm$0.22 & 0.23$\pm$0.34 & 0.67$\pm$0.27\\
4.0 & 0.07$\pm$0.29 & -0.32$\pm$0.28 & 0.42$\pm$0.32 & 0.88$\pm$0.29\\
4.3 & 0.12$\pm$0.3 & -0.21$\pm$0.29 & 0.5$\pm$0.32 & 0.98$\pm$0.29\\
4.6 & 0.17$\pm$0.31 & -0.09$\pm$0.3 & 0.59$\pm$0.31 & 1.09$\pm$0.29\\
5.0 & 0.24$\pm$0.31 & 0.03$\pm$0.31 & 0.7$\pm$0.32 & 1.21$\pm$0.26\\
5.5 & 0.31$\pm$0.32 & 0.16$\pm$0.32 & 0.79$\pm$0.3 & 1.32$\pm$0.26\\
5.9 & 0.41$\pm$0.33 & 0.29$\pm$0.33 & 0.88$\pm$0.29 & 1.47$\pm$0.24\\
6.4 & 0.51$\pm$0.34 & 0.43$\pm$0.34 & 0.98$\pm$0.29 & 1.68$\pm$0.12\\
7.0 & 0.63$\pm$0.36 & 0.57$\pm$0.36 & 1.07$\pm$0.28 & 1.68$\pm$1.68\\
7.6 & 0.76$\pm$0.38 & 0.72$\pm$0.38 & 1.15$\pm$0.28 & --\\
8.2 & 0.9$\pm$0.4 & 0.88$\pm$0.4 & 1.22$\pm$0.31 & --\\
8.9 & 1.05$\pm$0.42 & 1.03$\pm$0.42 & 1.32$\pm$0.35 & --\\
9.7 & 1.19$\pm$0.4 & 1.17$\pm$0.4 & -- & --\\
10.6 & 1.39$\pm$0.44 & 1.38$\pm$0.44 & -- & --\\
11.5 & 1.57$\pm$0.44 & 1.57$\pm$0.44 & -- & --\\
12.5 & 1.74$\pm$0.36 & 1.74$\pm$0.36 & -- & --\\
\hline
\end{tabular}
\tablefoot{\sigmasfr\ is given in units of \msun/yr/kpc$^2$. The uncertainties represent the $1\sigma$ dispersion of the points in a particular redshift and stellar mass bin.}
\end{center}
\end{table}


\subsection{SFR surface density main sequence}

\begin{figure*}[ht!]
\includegraphics[width=0.5\textwidth,clip]{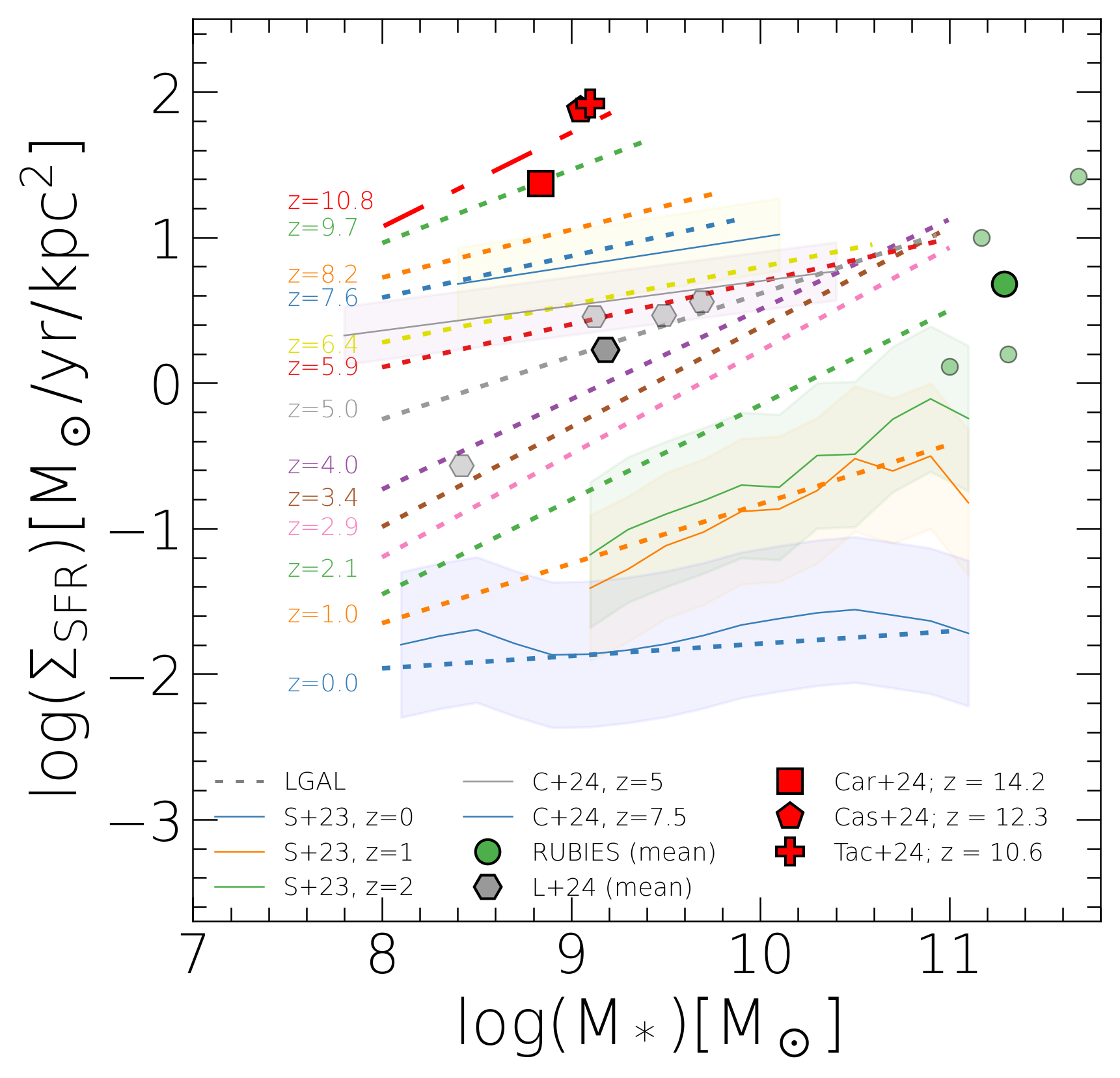}
\includegraphics[width=0.5\textwidth,clip]{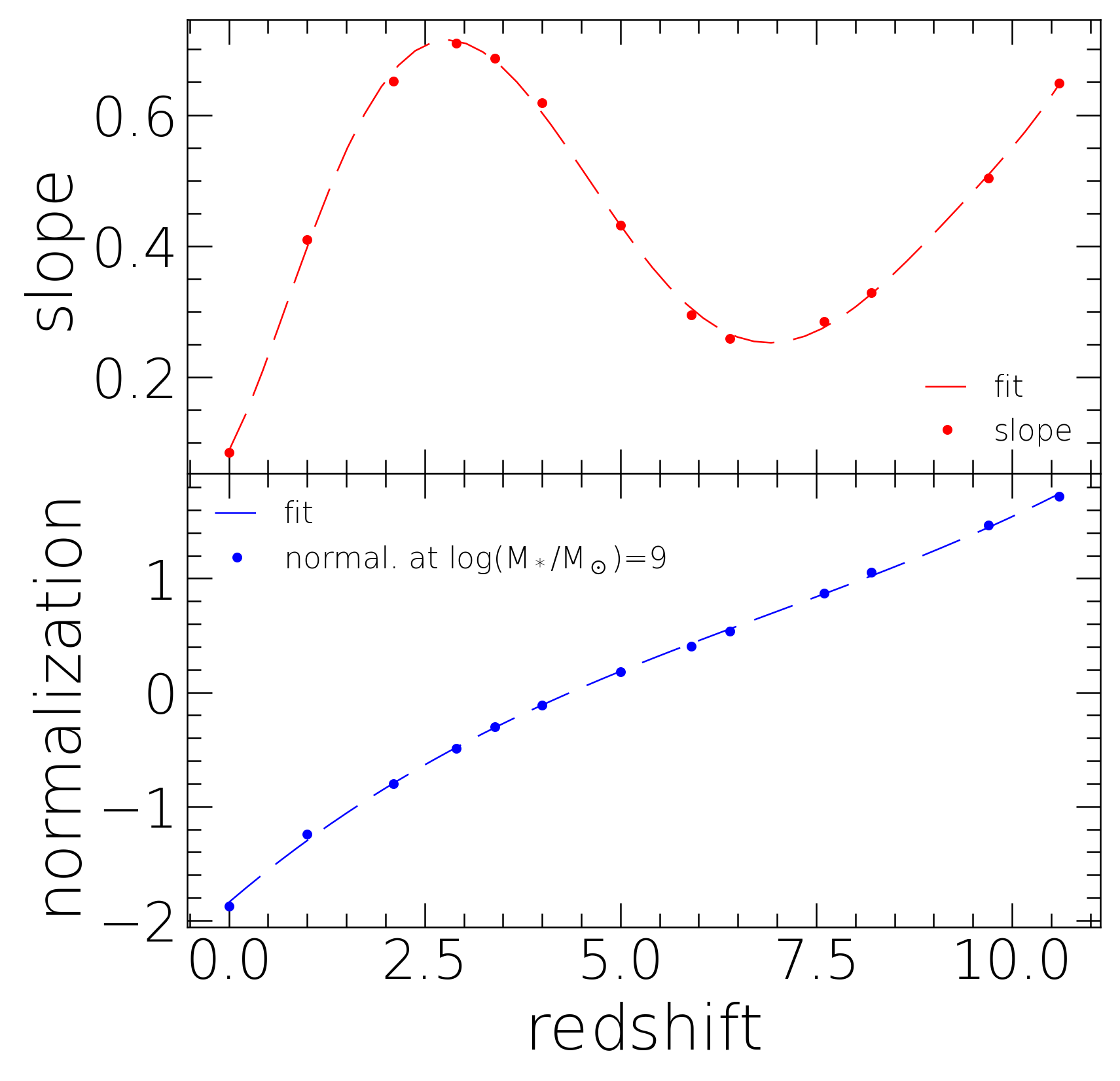}
\caption{Star formation surface density \sigmasfr\ main sequence (\sigmaMS). The dotted lines show \sigmaMS\ estimated using simulated galaxies from \lgalaxies\ at a given reshift from $z=0$ up to $z\sim10.8$ (the last dot-dashed fit includes three snapshots to increase the sample size; see Tab. \ref{tab:fittingSigmaSFR}). {  The colors indicate the (mean) redshift of the points corresponding to the \sigmaMS\ with the same color.} \citet{Salim2023ApJ...958..183S} \sigmaMS\ at $z=0, 1, 2$, and \citet{Calabro2024arXiv240217829C} at $z=5$ and $7.5$ are shown with solid lines in blue, orange, green, purple, and yellow with the corresponding scatter and within the stellar mass range studied in their work. The markers show individual intermediate- and high-$z$ galaxies. {  {\it Right panel:} Evolution of the slope and normalization of the estimated \sigmaMS. The polynomial coefficients are given in Equations \ref{eq:slope} and \ref{eq:ynorm}.}
\label{fig:SigmaSFR_MS}}
\end{figure*}

We focus below on the evolution of \sigmaMS\ from $z=11$ to $z=0$ in the stellar mass range as shown in Figure \ref{fig:SigmaSFR_MS}. The fit to the selected low-$z$ simulated galaxies agrees with \sigmaMS\ at $z=0$ and $1$ \citep{Salim2023ApJ...958..183S}. As shown in Figure \ref{fig:SigmaSFR_MS}, our \sigmaMS\ at $z=0$ and $1$ are within the 0.5 dex scatter of the \citet{Salim2023ApJ...958..183S} \sigmaMS\ with similar slopes. In the case of \sigmaMS\ at $z=2$, the slope from \citet{Salim2023ApJ...958..183S} is similar to the simulated \sigmaMS\ at $z=2$, but with a lower normalization by $\sim0.5$ dex. {Only 15\% of the $z=2$ sample from \citet{Salim2023ApJ...958..183S} have spectroscopic redshifts, which may increase the uncertainties of the estimated parameters. At the same redshift, we show four massive ($>10^{11}$\msun) dusty SFGs from the RUBIES survey \citep[][in green]{Cooper2024Rubies}.
These are above the \sigmaMS\ from \citet{Salim2023ApJ...958..183S}, but agree with the extrapolation of \sigmaMS\ derived in this work.}
The observed \sigmaMS\ at $z=5$ and $7.5$ from \citet{Calabro2024arXiv240217829C} both agree with the simulated \sigmaMS\ at the same redshifts, { except for the low-mass end at $z=5$, where the observed \sigmasfr\ are $\sim0.5$ dex higher. As noted in \citet{Calabro2024arXiv240217829C}, this may be caused by excluding the diffuse warm ionized gas and/or collisional extinction, which may contribute to the H$\alpha$ emission and lead to overestimating the SF. The data in \citet{Calabro2024arXiv240217829C} were moreover binned over a wide range of redshifts (4 to 6, and 6 to 10), which may also cause the observed difference. A smaller range would considerably reduce the number of sources in the narrower bins. Finally, we show individual galaxies at redshift $~4.5$ and above $10$. These are found around (within 0.4 dex) the simulated \sigmaMS\ at a given redshift.}

The general trend of an increasing normalization in \sigmaMS\ with redshift is visible in \lgalaxies\ (right panel in Fig. \ref{fig:SigmaSFR_MS}). This increase in normalization at the most basic level is due to the overall decrease in the galaxy sizes for increasing SFRs with redshift (Fig. \ref{fig:bt_size_redshift}, top right, and bottom left panels). Since the \sigmasfr\ is proportional to SFR and inversely proportional to the square of the size of a galaxy, changes in these two parameters work toward increasing \sigmasfr\ with redshift.

The \sigmaMS\ slope rises rapidly between $z=1$ and $3$. This rise was pointed out by \citet{Salim2023ApJ...958..183S}. They suggested that the bulge build-up in more massive galaxies (or being more compact) caused this increase in the slope, up to $z=2$ in their work. Here, the highest slope of 0.71 was found at $z=2.9$, which roughly corresponds to the peak of the CSFRD. As shown in Figure \ref{fig:SFRD_MD14}, for  $z < 4.5$, the most massive galaxies begin to dominate the CSFRD (see Appendix \ref{sec:csfrd_appx} for more details and a comparison with observations). At the same redshift, we also find that the most massive galaxies show the fastest evolution of \sigmasfr\ (Fig. \ref{fig:sigma_sfr}). All these pieces of evidence indicate that the most massive galaxies { (and in particular, the drastic drop in SFR)} cause the change in the slope of \sigmaMS. { Stated differently, the star-forming activity in most massive galaxies has dropped more than in less massive galaxies since $z\sim3$.} We discuss the possible cause of this slope evolution in the next section.

\begin{table}[ht]
\begin{center}
	\caption{Results of the \sigmaMS\ fitting.
 \label{tab:fittingSigmaSFR}}
\begin{tabular}{cccccr}
$z$  & {\it a}  & {\it b} & \logmass  & sample \\
 & & & range & size \\
\hline
0.0 & 0.085$\pm$0.003 & -1.88$\pm$0.01 & 8.0 -- 11.0 & 309977\\
1.0 & 0.41$\pm$0.003 & -1.24$\pm$0.01 & 8.0 -- 11.0 & 313879\\
2.1 & 0.651$\pm$0.004 & -0.8$\pm$0.01 & 8.0 -- 11.0 & 257737\\
2.9 & 0.709$\pm$0.005 & -0.49$\pm$0.01 & 8.0 -- 11.0 & 207934\\
3.4 & 0.686$\pm$0.006 & -0.30$\pm$0.01 & 8.0 -- 11.0 & 173087\\
4.0 & 0.619$\pm$0.007 & -0.11$\pm$0.01 & 8.0 -- 11.0 & 132988\\
5.0 & 0.432$\pm$0.011 & 0.18$\pm$0.01 & 8.0 -- 10.9 & 71266\\
5.9 & 0.295$\pm$0.018 & 0.4$\pm$0.01 & 8.0 -- 11.0 & 37196\\
6.4 & 0.259$\pm$0.023 & 0.54$\pm$0.01 & 8.0 -- 10.6 & 24324\\
7.6 & 0.285$\pm$0.041 & 0.87$\pm$0.03 & 8.0 -- 9.9 & 8074\\
8.2 & 0.329$\pm$0.06 & 1.05$\pm$0.04 & 8.0 -- 9.8 & 3938\\
9.7 & 0.504$\pm$0.154 & 1.46$\pm$0.1 & 8.0 -- 9.4 & 625\\
\hline
10.6 & 0.575$\pm$0.283 & 1.65$\pm$0.18 & 8.0 -- 9.2 & 171\\
11.5 & 0.95$\pm$0.655 & 1.97$\pm$0.37 & 8.0 -- 9.0 & 42\\
12.5 & 0.637$\pm$1.593 & 1.79$\pm$0.95 & 8.2 -- 8.8 & 7\\
\hline
10.8 & 0.65$\pm$0.26 & 1.72$\pm$0.16 & 8.0 -- 9.2 & 220\\
\end{tabular}
\tablefoot{{\it a} and {\it b} are the slope and intercept {  evaluated at log(\mstar/\msun)$=9$ of the fits in the form log(\sigmasfr)$=a\times [\log(M_*/\msun)-9] +b$} and shown in Figure \ref{fig:SigmaSFR_MS}. To increase the sample size for the last three snapshots, from $z=10.6$ to $12.5$, they were analyzed together, and the fit coefficients are given in the last row.}
\end{center}
\end{table}

{ Using the slope and normalization evolution with redshift (see the right panel of Fig. \ref{fig:SigmaSFR_MS}), we derived the redshift-dependent \sigmaMS\ evolution. First, using the polynomial function, we estimated the evolution of the slope and the normalization with redshift. For the slope, we obtained a sixth-degree polynomial, 
\begin{equation}\label{eq:slope}
\begin{split}
y_{\rm slope} = 3.8972\times10^{-5} \times z^{6} -0.001401 \times z^{5} + 0.01848\times z^{4} \\-0.1024 \times z^{3}+ 0.1688 \times z^{2} + 0.2272\times z + 0.0877, 
\end{split}
\end{equation}
and for the normalization {  [evaluated at log(\mstar/\msun)=9]}, we obtained a third-degree polynomial, 
\begin{equation}\label{eq:ynorm}
\begin{split}
y_{\rm norm.} = 0.002315  \times z^{3} -0.04812   \times z^{2} + 0.5879 \times z -1.8406,
\end{split}
\end{equation}
where $z$ is redshift. {  The slope and normalization evolution are shown in the right panel of Figure \ref{fig:SigmaSFR_MS}.}
Finally, using these, we derived the \sigmaMS\ as a function of stellar mass and redshift in the form
\begin{equation}\label{eq:sigmsfrms_evol}
\log(\Sigma_{SFR})=y_{slope}(z) \times [\log(M_*/\msun) -9] + y_{norm.}(z),
\end{equation}
where $y_{slope}$ and $y_{norm.}$ are the functions of redshift given in Equation  \ref{eq:slope} and \ref{eq:ynorm}. 
}


\subsection{The cause of the slope evolution}
\begin{figure*}[ht!]
\includegraphics[width=0.24\textwidth,clip]{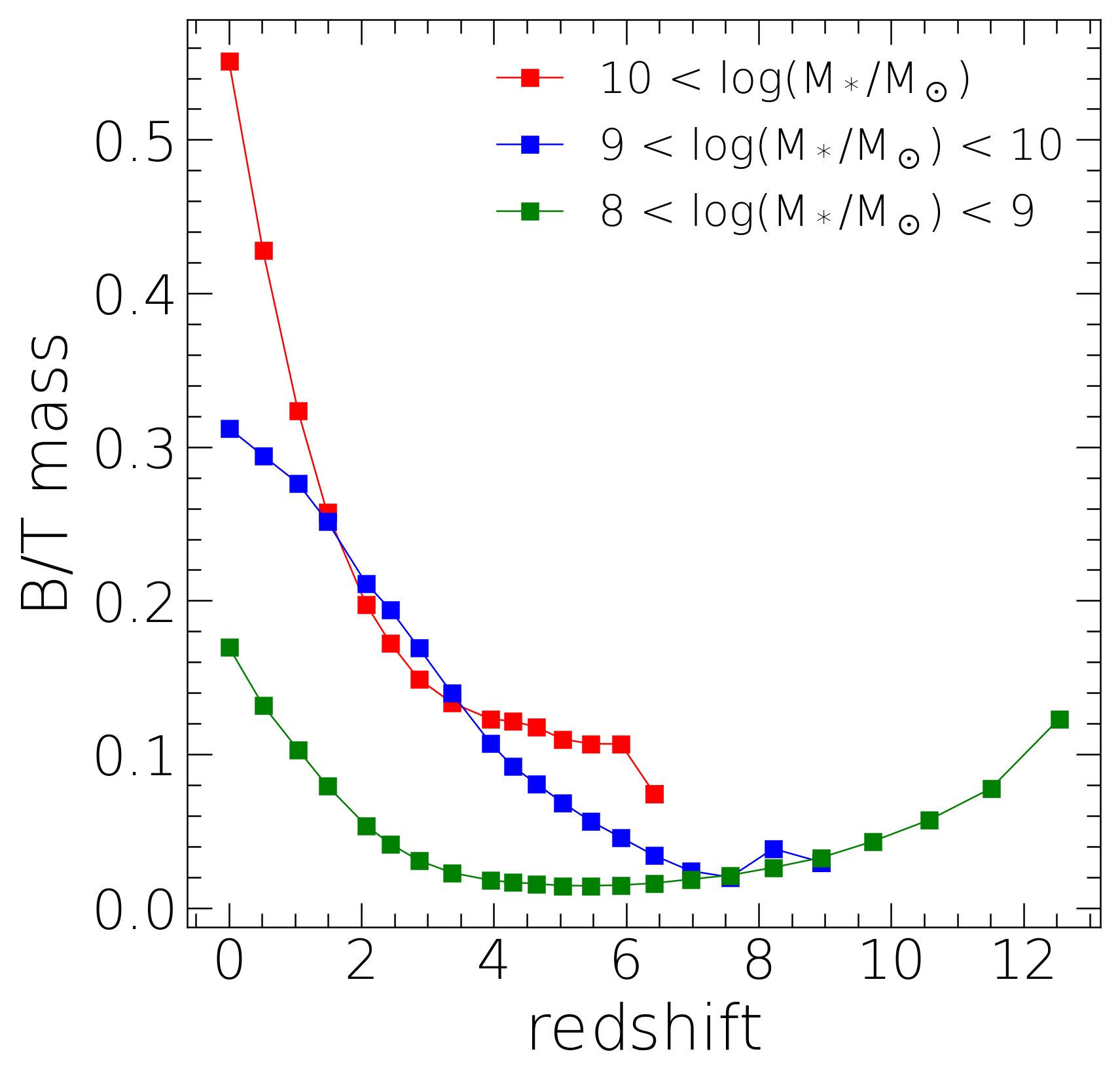}
\includegraphics[width=0.24\textwidth,clip]{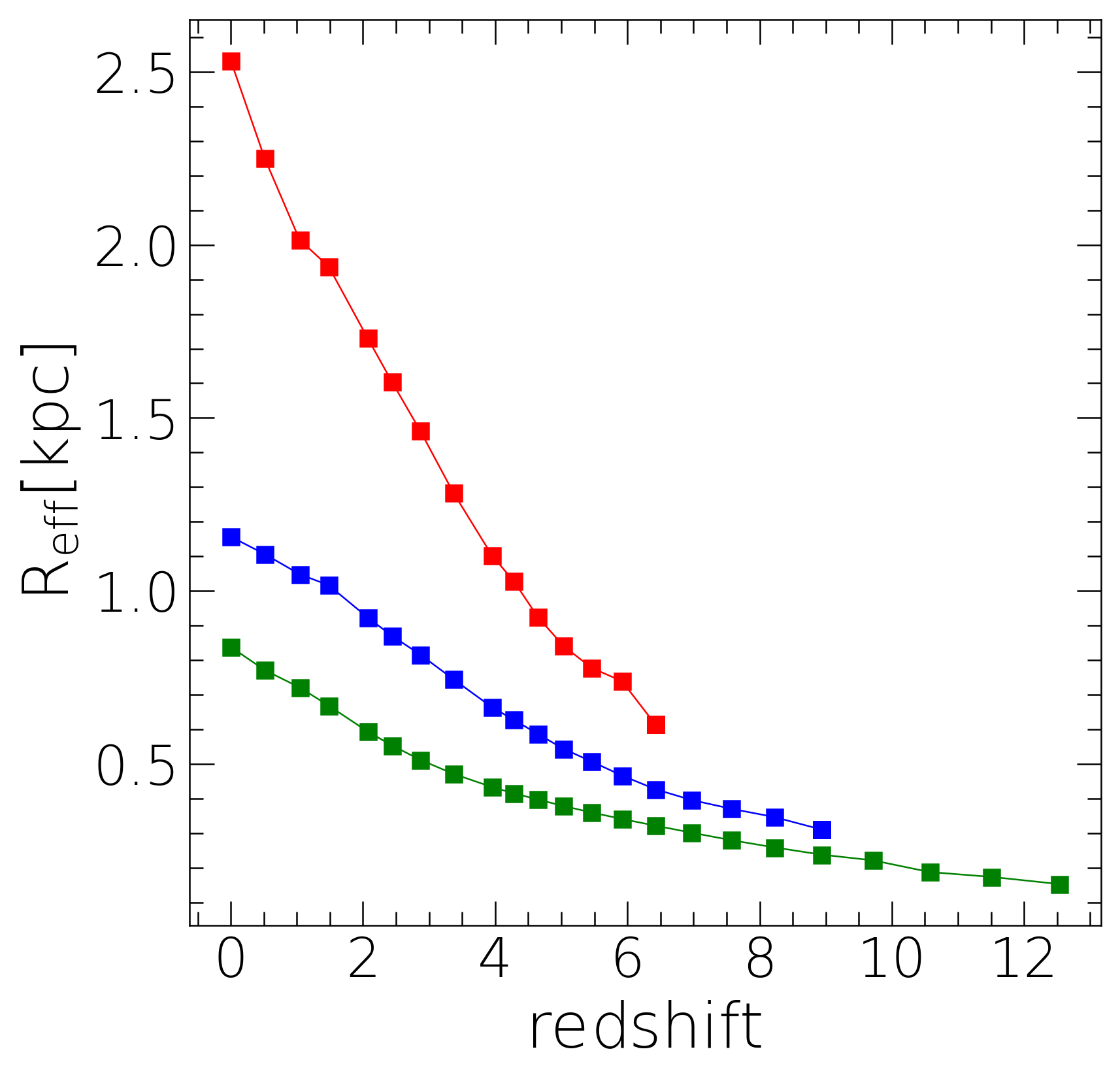}
\includegraphics[width=0.25\textwidth,clip]{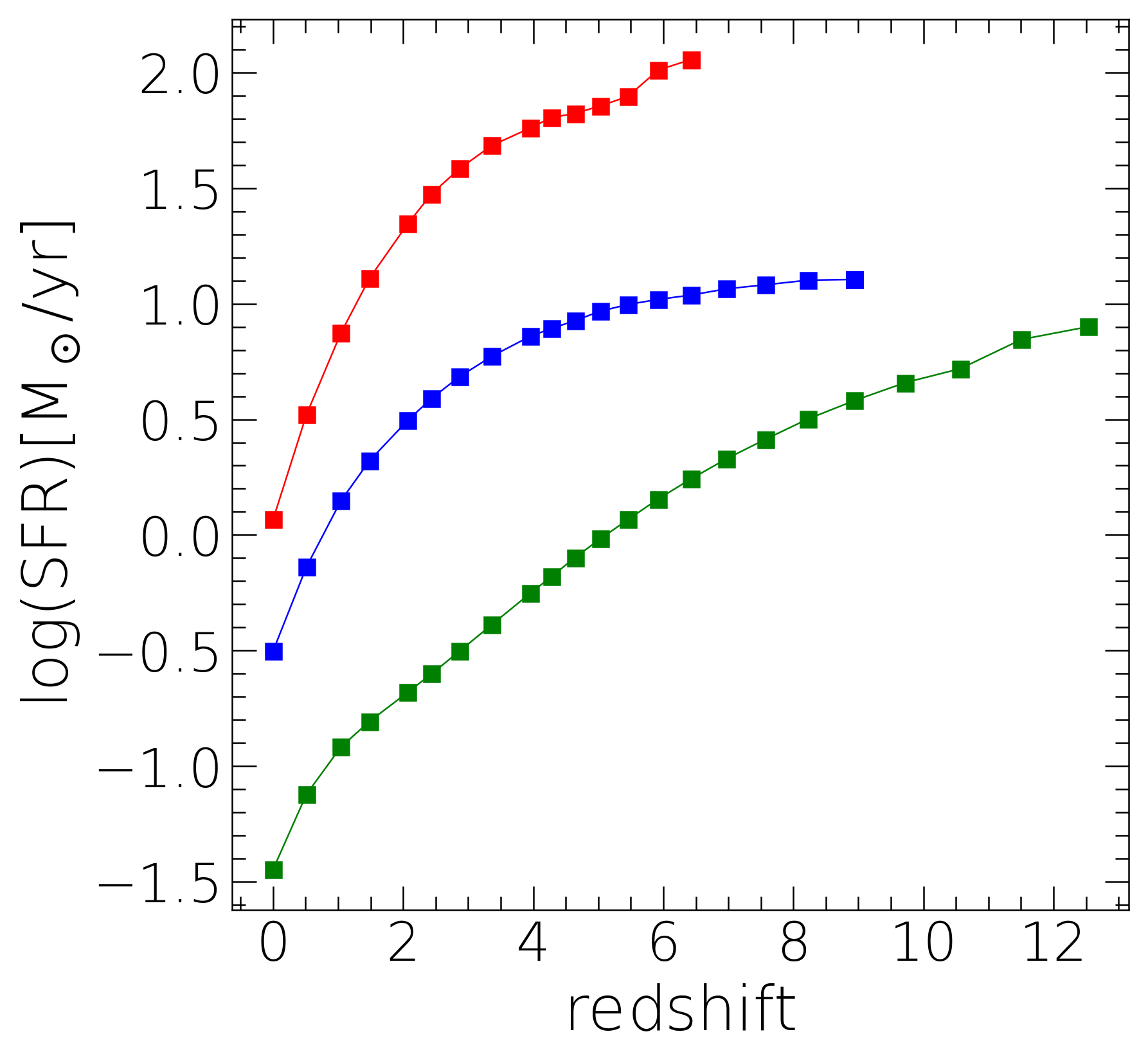}
\includegraphics[width=0.25\textwidth,clip]{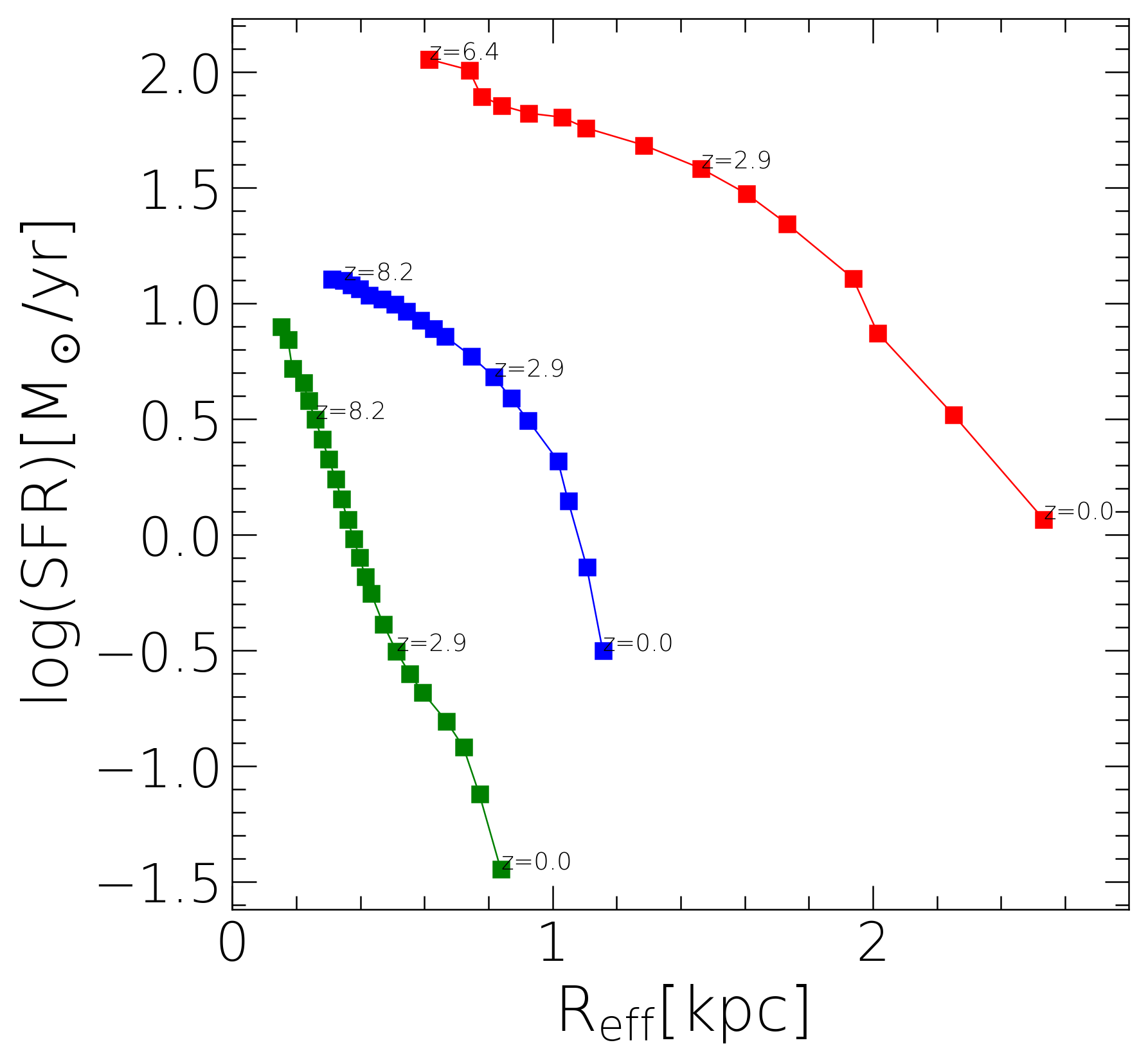}
\caption{Bulge-to-total mass ratio, effective radius, and SFR as a function of redshift, and SFR as a function of effective radius for simulated galaxies in three stellar mass bins as indicated in the legend. The redshifts in the right panel are shown for reference. 
\label{fig:bt_size_redshift}}
\end{figure*}

We investigated what might cause the slope decrease between $z=4$ and $1$. To verify whether the slope of the \sigmaMS\ changes due to the bulge build-up in massive galaxies (as suggested by \citealt{Salim2023ApJ...958..183S}), we considered the evolution of the bulge-to-total mass ratio (B/T) in stellar mass bins shown in the top left panel of Figure \ref{fig:bt_size_redshift}.  For massive and intermediate-mass galaxies, their B/T increases similarly between $z=4$ and $2$. Later, below $z=2$, the most massive galaxies build their bulges in a pronounced manner to $z=0$. The low-mass galaxies have a low ($<20\%$) bulge contribution to the total mass. 
The top right panel Fig. \ref{fig:bt_size_redshift} shows a distinct $R_e$ evolution for all three stellar mass bins in the same redshift range. The most massive galaxies increase their size by $\sim$200\% from $z=4$ to $z=0$, while the intermediate- and low-mass galaxies increase their sizes by $\sim$160\%\ in the same $z$ range.

We then considered the changes in SFR (bottom left panel Fig. \ref{fig:bt_size_redshift}). In the same redshift range, there is a drop of $\sim5000\%$ (1.9 dex), $\sim2000\%$ (1.3 dex), and $\sim1500\%$ (1.2 dex) for massive, intermediate-, and low-mass galaxies. Thus, the change in \sigmasfr\ mostly depends on the changes in SFR, at least since $z\sim4$, while the change in $R_e$ also increases \sigmasfr. The slope of \sigmaMS\ at $z>4$ is lower, and we can also observe a much slower SFR evolution in the most massive and intermediate-mass galaxies.

Thus, even if \sigmasfr\ is inversely proportional to the square of the size, the much greater changes in SFR (for the most massive galaxies in particular) drive the slope evolution with redshift between $z=4$ and $1$. {  The bulge build-up alone is not sufficient to explain the slope evolution.}

Considering the slope increase between $z=7.5$ and $3$ (Fig. \ref{fig:SigmaSFR_MS}), we find that the appearance of the most massive highly star-forming galaxies seems to cause this. These massive galaxies with SFRs 1\,dex higher than the lower-mass galaxies (at $z\sim7$) drag up the high-mass end of the \sigmaMS. This increase continues until the peak of the CSFRD at $z\sim2.5$.


\section{Discussion and conclusions}
\label{sec:discussion}
We showed the evolution of \sigmasfr\ and its main sequences \sigmaMS\ since $z=12$ to $z=0$. This sets the evolutionary sequences for this physically motivated star formation indicator.

Considering the whole \lgalaxies\ population in snapshots from $z =12$ to $z=0$, we find the evolution of \sigmasfr\ with redshift by  $\sim3$ dex (considering the mean values of the whole population; black squares in Fig. \ref{fig:sigma_sfr}). 

The observational data from JWST at $4 < z < 10$ \citep{Calabro2024arXiv240217829C} agree with the simulated population with stellar masses below $10^{10}$\msun. In particular, their fit runs parallel and between our low- and intermediate-mass bins with \mstar$<10^{10}$\msun. A decrease in \sigmasfr\ from $\sim25$ to $\sim1$ \msun/yr/kpc$^2$ is observed from $z=10$ to $4$ (see Fig. \ref{fig:sigma_sfr}). This also agrees with previous investigations of the \sigmasfr\ evolution in this redshift range using simulated CEERS light cones \citep{Yung2022_SC_SAM}. The most massive galaxies have higher \sigmasfr\ at virtually all redshifts. Furthermore, these objects show the fastest evolution of \sigmasfr\ from $z=4$ to $z=0$ while dominating the CSFRD in the same redshift range (see Fig. \ref{fig:SFRD_MD14}).

The simulated \sigmaMS\ agree with the observed relations at $z\sim0,1,7.5$. Considering the $z=2$ \sigmaMS, the simulated MS have a similar slope but a 0.5 dex higher normalization than is observed. This difference may be due to a low number of objects with spectroscopic redshifts for the observed $z=2$ sample \citep{Salim2023ApJ...958..183S}. However, massive dusty star-forming galaxies from the RUBIES survey at $z\sim2$ follow the simulated \sigmaMS\ at this redshift. For the \sigmaMS\ at $z=5$, the simulated MS is steeper, and the low-mass end lies below the observed relation. 

In the context of \sigmasfr\ and its main sequences, we showed very distant ($z>10$) spectroscopically confirmed galaxies. These follow (roughly) the evolution of \sigmasfr\ considering their stellar mass, and they are close to the predicted simulated \sigmaMS\ at a similar redshift. This shows the SAM success in predicting such high-$z$ \sigmasfr.

We described the evolution of simulated \sigmaMS\ from $z=12$ to $z=0$ (see Fig. \ref{fig:SigmaSFR_MS}, and Tab. \ref{tab:fittingSigmaSFR}). 
The slope of the \sigmaMS\ decreases from 0.71 at $z=2.9$ to 0.08 at $z=0$. We found that this decrease in the slope is due to a dramatic decrease (by $\sim5000\%$) in SFR in this redshift range for the most massive galaxies. This is a dominant factor considering that the changes in $R_e$ are much smaller.

Finally, using the slope and normalization evolution (Eq. \ref{eq:slope} and \ref{eq:ynorm}, and the right panel of Fig. \ref{fig:SigmaSFR_MS}), we provided the evolution of \sigmaMS\ as a function of stellar mass and redshift within the SAM framework (Eq. \ref{eq:sigmsfrms_evol}), up to $z=10.8$ with the stellar mass ranges indicated in Table \ref{tab:fittingSigmaSFR}.


\begin{acknowledgements}
J.N., M.J.M., M.S. and A.L.~acknowledge the support of 
the National Science Centre, Poland through the SONATA BIS grant 2018/30/E/ST9/00208.
This research was funded in whole or in part by National Science Centre, Poland (grant numbers: 2021/41/N/ST9/02662
and 2023/49/B/ST9/00066). 
J.N.~acknowledges the support of the Polish National Agency for Academic Exchange (NAWA) Bekker  grant BPN/BEK/2023/1/00271, and the kind hospitality of the IAC. M.J.M.~acknowledges the support of the NAWA Bekker grant BPN/BEK/2022/1/00110. 

Authors acknowledge the use of astropy libraries \citep{2013A&A...558A..33A,2018AJ....156..123A}, as well as TOPCAT/STILTS software \citep{Taylor2005ASPC..347...29T}.
\end{acknowledgements}

\bibliographystyle{aa} 
\bibliography{biblography}

\begin{appendix}


\section{Cosmic star-formation rate density}
\label{sec:csfrd_appx}
\begin{figure}[ht!]
\includegraphics[width=0.5\textwidth,clip]{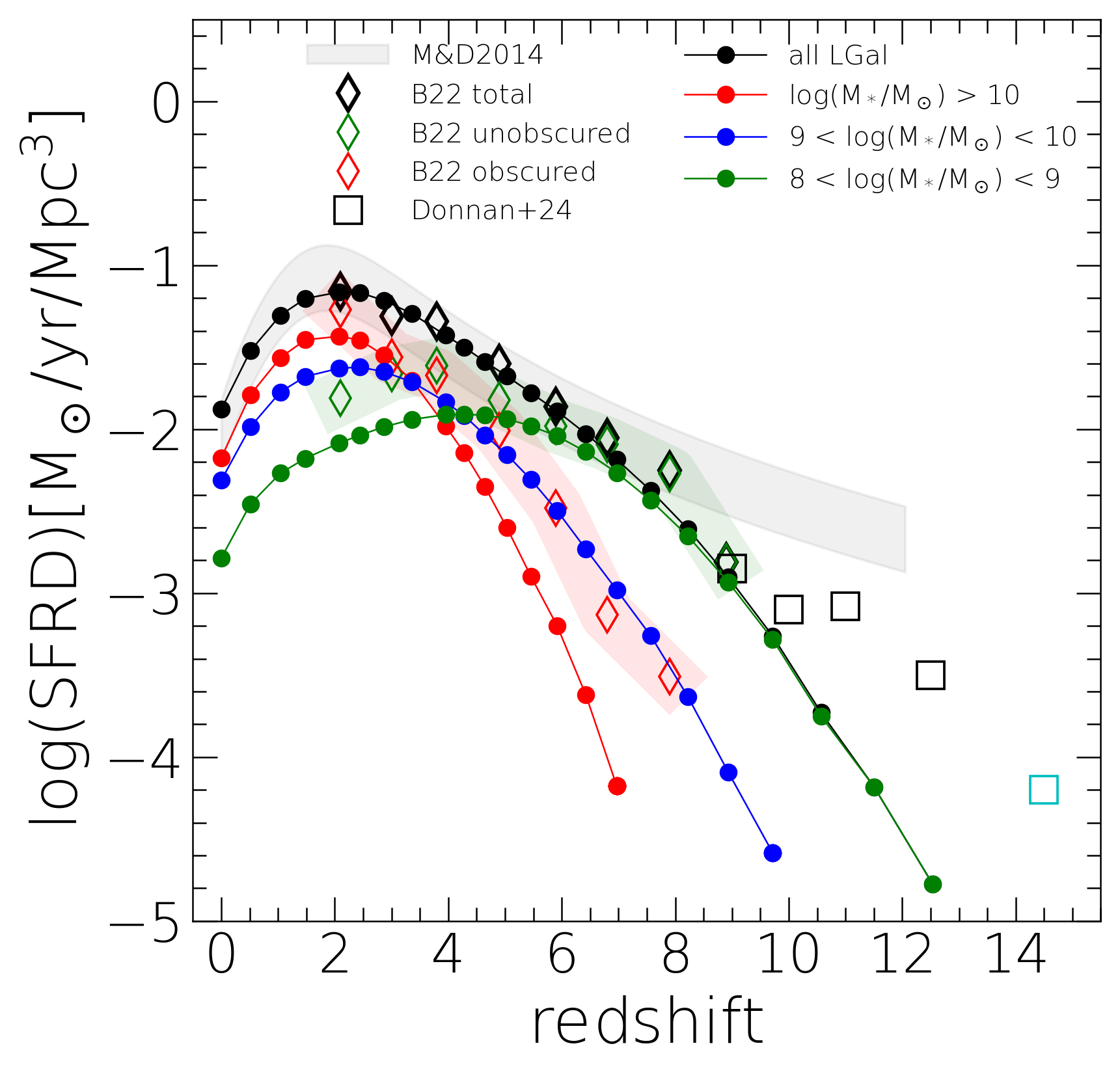}
\caption{Cosmic star-formation rate density. Gray shaded region shows CSFRD from \citet{MadauDickinson2014} scaled to the \citet{Chabrier2003PASP..115..763C} IMF. Black dots and line represent a total CSFRD estimated using \lgalaxies\ above the stellar mass limit ($>10^8$\msun). The red, blue and green points and lines represent \lgalaxies\ subsamples in stellar mass bins. Diamonds show data from \citet{Bouwens_2022} for total (black), unobscured (green) and obscured (red) estimates of CSFRD, with shaded regions only for visualization purposes; squares show data from \citet{Donnan2024MNRAS}, while the cyan square is the tentative estimates based on a scarce sample size at $z=14.5$ from their work.
\label{fig:SFRD_MD14}}
\end{figure}

The cosmic star-formation rate density (CSFRD) is shown in Figure \ref{fig:SFRD_MD14}. The total simulated CSFRD has been shown to reproduce observations \citep[][]{Parente2023MNRAS.tmp..881P}. Here we show the total CSFRD up to $z=12$ together with the CSFRD for the three stellar mass bins. It is clear that at higher-$z$ the low-mass galaxies dominate the cosmic SFRD, while at $z\sim 4$, the most massive galaxies take over the low and mid-mass galaxies leading to the peak of CSFRD at $z\sim 2$. \citet{Bouwens_2022} studied obscured and unobscured CSFRD at redshift range from $2$ up to $9$. They showed that obscured CSFRD began to dominate the total CSFRD at $z<4$ \citep[see also][]{Zavala2021ApJ}, coinciding with the dominance of massive galaxies from SAM. Above that $z$, the obscured CSFRD drops following the mid-mass galaxies, while the unobscured CSFRD follows a milder decline, as shown by low-mass galaxies. 

The total simulated CSFRD agrees with the observed SFRD from \citet[][scaled to the \citealt{Chabrier2003PASP..115..763C} IMF]{MadauDickinson2014} up to $z=8$. At $z>8$ the total simulated CSFRD shows lower values compared to \citet[][already extrapolated above $z\sim8$]{MadauDickinson2014}. The observed high-$z$ data points from \citet{Donnan2024MNRAS} show a similar drop in CSFRD below \citet{MadauDickinson2014} fit. As suggested by \citet{Donnan2024MNRAS}, there is tentative evidence for CSFRD rapid decline at $z>13$.

\end{appendix}
\end{document}